\documentclass[twocolumn,journal]{IEEEtran}
\usepackage{amsfonts}
\usepackage{amssymb}
\usepackage{amsmath}
\usepackage{dsfont}
\usepackage{graphicx,subfigure}
\usepackage{enumitem}
\usepackage{cite}
\usepackage{bigstrut}
\usepackage{float}
\usepackage{balance}
\usepackage{lipsum}
\usepackage{psfrag}
\usepackage{algorithm}
\usepackage{color}
\usepackage{booktabs}
\usepackage[noend]{algpseudocode}
\usepackage[export]{adjustbox}

\newtheorem{proposition}{Proposition}

\usepackage{array}
\newcolumntype{P}[1]{>{\centering\arraybackslash}p{#1}}
 
\begin{document}

\title{StatAvg: Mitigating Data Heterogeneity in Federated Learning for Intrusion Detection Systems}

\author{Pavlos~S.~Bouzinis\IEEEauthorrefmark{2}, Panagiotis Radoglou-Grammatikis\IEEEauthorrefmark{3}\IEEEauthorrefmark{4}, Ioannis Makris\IEEEauthorrefmark{2}, Thomas Lagkas\IEEEauthorrefmark{5}, Vasileios Argyriou\IEEEauthorrefmark{6}, Georgios~Th.~Papadopoulos\IEEEauthorrefmark{7}, Panagiotis Sarigiannidis\IEEEauthorrefmark{3} and George~K.~Karagiannidis\IEEEauthorrefmark{8}\IEEEauthorrefmark{9}

\thanks{\IEEEauthorrefmark{1}\textit{This project has received funding from the European Union’s Horizon Europe research and innovation programme under grant agreement No 101070450 (\texttt{AI4CYBER}). Disclaimer: Funded by the European Union. Views and opinions expressed are, however, those of the author(s) only and do not necessarily reflect those of the European Union or European Commission. Neither the European Union nor the European Commission can be held responsible for them.}}
\thanks{\IEEEauthorrefmark{2}P.~S.~Bouzinis and I.~Makris are with MetaMind Innovation P.C., Kila Kozani, 50100, Kozani, Greece - \texttt{E-Mail: pbouzinis@metamind.gr; makris@metamind.gr}}
\thanks{\IEEEauthorrefmark{3}P.~Radoglou-Grammatikis and P.~Sarigiannidis are with the Department of Electrical and Computer Engineering, University of Western Macedonia, Campus ZEP Kozani, 50100, Kozani, Greece - \texttt{E-Mail: pradoglou@uowm.gr; psarigiannidis@uowm.gr}}
\thanks{\IEEEauthorrefmark{4}P.~Radoglou-Grammatikis is also with K3Y Ltd, William Gladstone 31, 1000, Sofia, Bulgaria - \texttt{E-Mail: pradoglou@k3y.bg}}
\thanks{\IEEEauthorrefmark{5}T.~Lagkas is with the Department of Computer Science, Democritus University of Thrace, Kavala Campus, 65404, Kavala, Greece - \texttt{E-Mail: tlagkas@cs.duth.gr}}
\thanks{\IEEEauthorrefmark{6}V.~Argyriou is with the Department of Networks and Digital Media, Kingston University London, Penrhyn Road, Kingston upon Thames, Surrey KT1 2EE, London, UK - \texttt{E-Mail: vasileios.argyriou@kingston.ac.uk}}
\thanks{\IEEEauthorrefmark{7}G.~Th.~Papadopoulos is with the Department of Informatics and Telematics, Harokopio University of Athens, Omirou 9, Tavros, GR17778, Athens, Greece - \texttt{E-Mail: g.th.papadopoulos@hua.gr}}
\thanks{\IEEEauthorrefmark{8}G. K. Karagiannidis is with the Department of Electrical and Computer Engineering, Aristotle University of Thessaloniki, 54124, Thessaloniki, Greece - \texttt{E-Mails: geokarag@auth.gr}}
\thanks{\IEEEauthorrefmark{9}G. K. Karagiannidis is also with the Artificial Intelligence \& Cyber Systems Research Center, Lebanese American University (LAU), 1102 2801 Beirut, Lebanon}

} 
\vspace{-0cm}
 \maketitle

\begin{abstract}
Federated learning (FL) is a decentralised learning technique that enables participating devices to collaboratively build a shared Machine Learning (ML) or Deep Learning (DL) model without revealing their raw data to a third party. Due to its privacy-preserving nature, FL has sparked widespread attention for building Intrusion Detection Systems (IDS) within the realm of cybersecurity. However, the data heterogeneity across participating domains and entities presents significant challenges for the reliable implementation of an FL-based IDS. In this paper, we propose an effective method called Statistical Averaging (StatAvg) to alleviate non-independently and identically (non-iid) distributed features across local clients' data in FL. In particular, StatAvg allows the FL clients to share their individual data statistics with the server, which then aggregates this information to produce global statistics. The latter are shared with the clients and used for universal data normalisation. It is worth mentioning that StatAvg can seamlessly integrate with any FL aggregation strategy, as it occurs before the actual FL training process. The proposed method is evaluated against baseline approaches using datasets for network and host Artificial Intelligence (AI)-powered IDS. The experimental results demonstrate the efficiency of StatAvg in mitigating non-iid feature distributions across the FL clients compared to the baseline methods.
\end{abstract}


\section{Introduction}
\label{Introduction}

In the dynamic era of smart technologies, including the Internet of Things (IoT) \cite{deng2023trusted}, Artificial Intelligence (AI) \cite{taddeo2019trusting} and 6G ultra-dense wireless networks \cite{deng2023review, radoglou2022strategic}, the attack surface increases significantly. In particular, from single-step attacks, the attackers now have the ability to design and execute multi-step attack scenarios, targeting multiple systems and domains in a coordinated and synchronised manner. Some characteristic attack campaigns, according to MITRE ATT\&CK, are (a) C0034 - 2022 Ukraine Electric Power Attack and (b) C0022 - Operation Dream Job. Moreover, the rapid evolution of AI allows cyberattackers to create sophisticated cyberattacks that can adjust to mitigation activities and relevant countermeasures. Generative AI agents are able to automatically exploit one-day vulnerabilities. Finally, despite the fact that AI can play a significant role in defensive mechanisms, adversarial attacks \cite{asimopoulos2023breaching} can mislead AI-enabled decision-making processes. Therefore, it is evident that the presence of reliable and efficient intrusion detection mechanisms is more necessary than ever.

Traditionally, Intrusion Detection Systems (IDS) rely on signature-based methods, where predefined attack rules or patterns, referred to as signatures, are identified and compared with the monitoring data, thus alerting a potential threat if a match is found. For instance, \texttt{Snort} and \texttt{Suricata} are popular IDS in this category. On the other hand, in recent years, both Machine Learning (ML) and Deep Learning (DL) models have already demonstrated significant promise as a means to detect cyberattacks \cite{radoglou2021self}. However, it is worth mentioning that these models need the presence of appropriate security datasets that are often not publicly available, especially for critical domains \cite{radoglou2021self}. In addition, appropriate adjustments are required to re-train and integrate these models. Finally, conventional ML/DL methods are conducted in a centralised fashion, where a central entity collects all the necessary data from endpoints to construct training datasets and afterwards generates the ML/DL models. Although this approach successfully enables the detection of intrusions, it raises privacy concerns since endpoints' private data are shared with third parties. 

\begin{figure*}[httb!]
 \centering \includegraphics[width=0.75\linewidth]{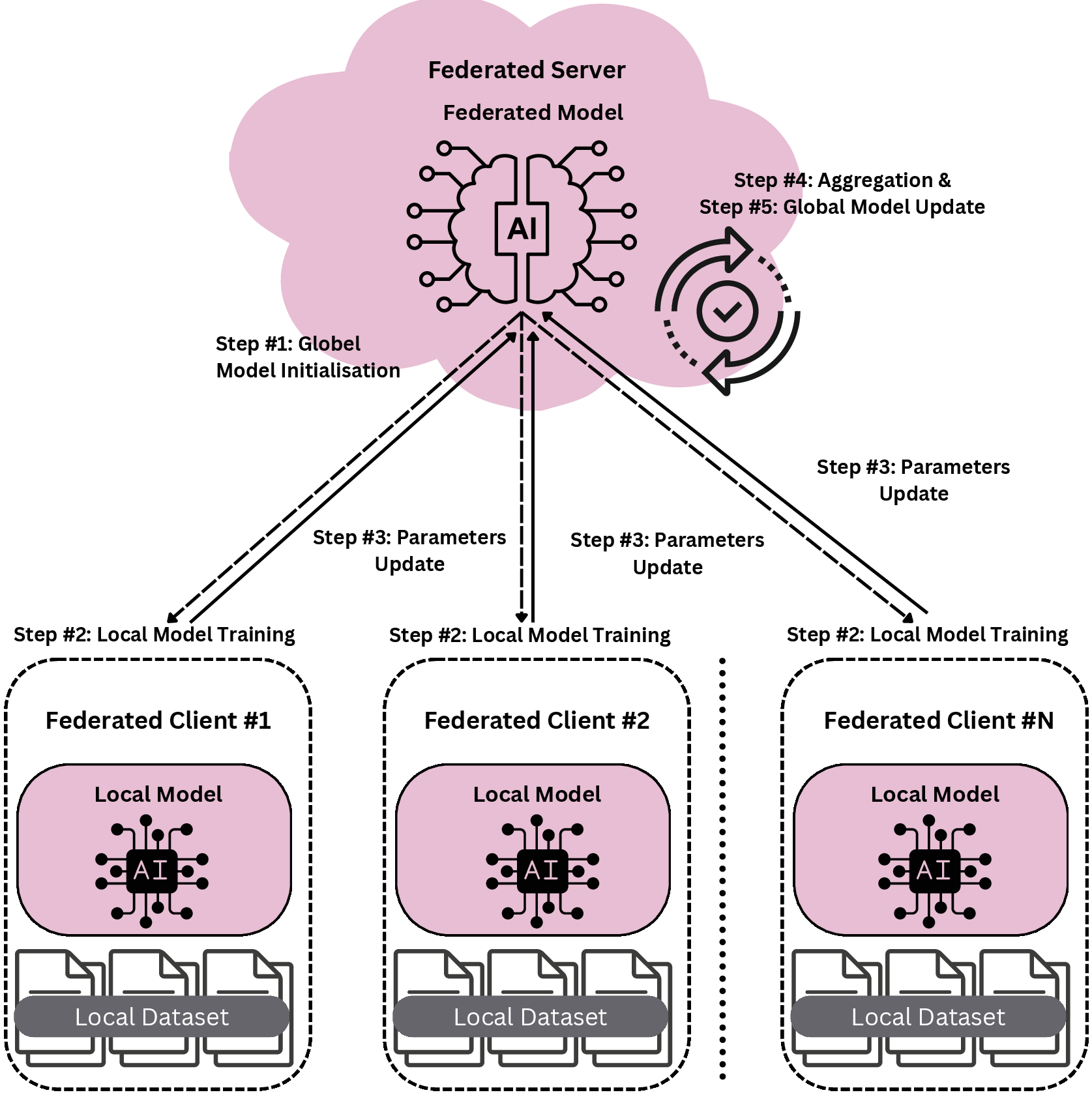}
 \caption{Federated Learning workflow.}
 \label{Federated Learning Workflow}
\end{figure*}

To alleviate privacy issues and mitigate communication overhead, Federated Learning (FL) has been proposed as an inherently privacy-preserving decentralised learning solution \cite{mcmahan2017communication, lin2023drl}. According to the FL principles, the participating clients are building an ML/DL model collaboratively with the aid of a central entity (e.g., a central server). The salient feature of FL is that clients transmit locally trained models to the server rather than raw data. Afterwards, the server aggregates the received parameters, updates the global model, and subsequently broadcasts it to the clients. As depicted in Fig.~\ref{Federated Learning Workflow}, this workflow is repeated until convergence is achieved. Consequently, the server has no access to clients' raw datasets. However, despite the benefits of FL, a notable challenge in the design of an FL-based IDS is the existence of non-independently and identically distributed (iid) data among clients, commonly referred to as data heterogeneity. In particular, if the data (including both benign and attack patterns) is not representative across the clients, the global model may become biased, thus working efficiently on some cases but inaccurately on others. Next, the presence of non-iid data can affect the federated training procedure in terms of delaying convergence. Finally, communication issues may also arise, given that more communication rounds might be required to ensure adequate model updates across the FL clients.

In light of the aforementioned remarks, in this paper, we introduce the \texttt{Statistical Averaging (StatAvg)} method to circumvent the challenges of non-iid features of clients in FL. Due to different feature distributions across clients, the local data normalisation process may differ from client to client. This inconsistency may slow down or even prevent the convergence of the federated global model since each local model is trained on a different input distribution. To this end, \texttt{StatAvg} aims at producing global data statistics that can serve as a universal normalisation for the local data of each client. To achieve this, the server is responsible for collecting the local statistics of the clients and afterwards aggregating them properly to produce global data statistics. In this manner, clients normalise their local data by sharing a common normalisation scaling, reflected by the global statistics, aimed at mitigating the effects of non-iid features present in their local datasets.

Therefore, the structure of this paper is organised as follows. Section~\ref{Related Work, Motivation and Contributions} discusses similar works in this field, thus drawing the motivation behind our work and highlighting our contributions. Section~\ref{Preliminaries of Federated Learning} provides preliminary information regarding FL. Next, section~\ref{StatAvg - Statistical Averaging} presents and analyzes \texttt{StatAvg}. Finally, section~\ref{Evaluation Analysis} focuses on the evaluation analysis of \texttt{StatAvg}, while section~\ref{Conclusions} concludes this paper.

\section{Related Work, Motivation and Contributions}
\label{Related Work, Motivation and Contributions}

\subsection{Related Work}
\label{Related Work}

Several works investigate the role and impact of FL in cybersecurity and, more precisely, in the scope of intrusion detection. Some survey papers in this field are listed in \cite{agrawal2022federated, campos2022evaluating, ghimire2022recent, arisdakessian2022survey, mothukuri2021survey}. In \cite{agrawal2022federated}, S. Agrawal et al. present a comprehensive survey regarding the impact of FL within the scope of intrusion detection, highlighting challenges and future directions. In \cite{campos2022evaluating}, E. Campos et al. provide a detailed comparison regarding (a) centralised, (b) distributed and (c) FL-driven intrusion detection mechanisms for IoT environments. Similarly, the authors in \cite{ghimire2022recent} discuss advances of FL within cybersecurity applications in IoT ecosystems. In \cite{arisdakessian2022survey}, S. Arisdakessian et al. provide a comprehensive study regarding FL-driven intrusion detection, game theory, social psychology and explainable AI. Finally, in \cite{mothukuri2021survey}, the authors focus their attention on security and privacy issues regarding FL applications. Next, we further discuss recent works that deliver FL-driven IDS, taking into account data heterogeneity. 

In \cite{li2020deepfed}, the authors introduce \texttt{DeepFed}, an FL-driven IDS for cyber-physical systems. The architectural design of \texttt{DeepFed} relies on three main elements: (a) Trust Authority, (b) Cloud Server and (c) $k$ Industrial Agents. The role of the Trust Authority is to produce the encryption keys for the proposed Pailier public-key cryptosystem utilised for the communication between the Cloud Server and the Industrial Agents. Next, the Cloud Server is responsible for the aggregation process, while the industrial agents focus on the local training procedure. Four attacks are taken into consideration by the proposed FL-driven IDS, namely (a) Reconnaissance Attacks, (b) Response Injection Attacks, (c) Command Injection Attacks and (d) Denial of Service (DoS) attacks. Moreover, two attacks targeting the FL process are considered: (a) Eavesdropping of Data Resources and (b) Eavesdropping of Model Parameters. The workflow of \texttt{DeepFed} follows five phases: (a) System Initialisation, (b) Local Model Training by Industrial Agents, (c) Model Parameters Encryption by Industrial Agents, (d) Model Parameters Aggregation by the Cloud Server and (e) Local Model Updating by Industrial Agents. For the detection process, the authors leverage a combined Convolutional Neural Network (CNN) - Gated Recurrent Unit (GRU), while special attention is paid to the proposed Pailier-based secure communication protocol for the communication between the Cloud Server and the Industrial Agents. Finally, three evaluation metrics are considered, namely Accuracy, Precision, Recall and F-Score, demonstrating the detection efficiency of \texttt{DeepFed}. 

In \cite{attota2021ensemble}, D. C. Attota et al. describe \texttt{MV-FLID}, a multi-view FL-based IDS which focuses on the detection of attacks against Message Queuing Telemetry Transport (MQTT) communications within IoT environments. In particular, MV-FLID adopts a multi-view approach, combining (a) bi-directional flow features (Bi-Flow View), (b) un-directional flow features (Uniflow View) and (c) packet features (Packet View). An FL model is generated for each of the previous viewpoints. Regarding the feature selection process, the authors leverage the Grey Wolf Optimizer (GWO) introduced by S. Mirjalili et al. in \cite{mirjalili2014grey}. Next, the federated training procedure follows, as depicted in Fig.~\ref{Federated Learning Workflow}. Finally, an ensemble-based technique is used to combine the outcomes of the FL models in order to provide a unified prediction outcome. Regarding the evaluation process, the authors use four metrics: (a) Accuracy, (b) F1 Score, (c) Precision and (d) Recall, thus demonstrating the overall detection effectiveness of \texttt{MV-FLID}.

In \cite{zhao2022semi}, R. Zhao et al. provide a semisupervised FL scheme for intrusion detection within IoT environments. The proposed scheme relies on CNN models, while four phases are followed in an iterative manner within the FL fashion: (a) client training, (b) knowledge distillation, (c) discrimination between familiar and unfamiliar traffic packets and (d) hard-labelling and voting. During the first phase, the clients train their CNNs with private local data. In the second phase, knowledge distillation follows a teacher-student approach, where a teacher model guides the training of a student model, providing soft targets or logits. In this case, the FL server serves as the teacher, aggregating the logits of each class from all clients and broadcasting them to each client. Then, the clients use the global logits as soft targets in order to train their local models, leveraging the distillation loss between the predicted logits and the global logits. Next, a discrimination network is used from the FL server to evaluate further the predicted labels of each client's CNN and provide feedback regarding the quality of the labels. Finally, hard labelling and voting mechanisms take place in order to consider only the labels from the majority of the FL clients and proceed with the aggregation process.

In \cite{chen2020intrusion}, the authors propose the FL-based Attention-Gated Recurrent Unit \texttt{(FedAGRU)} to address, among others, the issue of different label distribution across clients' data, thus demonstrating performance gains over conventional FL aggregation strategies. Moreover, in \cite{wang2021non}, H. Wang et al. propose a peer-to-peer algorithm, namely P2PK-SMOTE, to train FL-driven anomaly detection models in non-iid scenarios. The latter refers to inter and intra-imbalanced classes across the FL clients. Additionally, S. Popoola et al. in \cite{popoola2021federated} use the \texttt{Fed+} method \cite{kundu2022robustness} for FL-driven intrusion detection in heterogeneous networks. The clients own datasets from various types of networks, such as industrial IoT, wireless networks and wireless vehicular networks, while \texttt{Fed+} facilitates the generation of personalised local models with enhanced attack classification performance. In \cite{ruzafa2021intrusion}, the authors take \texttt{Fed+} a step further by incorporating differential privacy techniques. Next, in \cite{weinger2022}, B. Weinger et al. investigate data augmentation techniques to address class imbalance and non-iid settings. The evaluation results indicate a performance improvement compared with baselines that do not rely on data augmentation strategies. Finally, in \cite{han2023heterogeneous}, H. Weixiang propose a clustering-enabled FL meta-learning framework to tackle class imbalance and non-iid data.

\subsection{Motivation}
\label{Motivation}

Undoubtedly, the previous works offer valuable insights and methodologies. However, in the majority of them, the assumption of iid across the clients is not valid within realistic FL conditions. Conventional FL strategies like FedAvg are not designed for handling non-iid data and may experience notable performance degradation or even divergence when applied in such situations \cite{kairouz2021}. Although the works \cite{chen2020intrusion, wang2021non} and \cite{ weinger2022, han2023heterogeneous, zhao2022semi} successfully examine and design FL-based IDS considering non-iid data, emphasis was mainly given to the following cases: (a) class imbalance across clients datasets and/or different label distributions and (b) different number of samples per client. On the contrary, the authors in \cite{popoola2021federated} study a broader aspect of non-iid settings by considering heterogeneous datasets across the clients. However, personalised FL methods such as Fed+ are employed, generating multiple personalised local models instead of a unified global one.

A particular example of non-iid data among clients is the case of non-iid features, which has generally received less attention in the FL-related literature. Methods addressing this issue mainly rely on layer normalisation \cite{du2022} and batch normalisation \cite{li2021fedbn} techniques. Specifically, \cite{li2021fedbn} proposes \texttt{FedBN}, a method that incorporates batch normalisation layers on local clients' model, which are not included in the aggregation step at the server side. Although \texttt{FedBN} has shown potential in mitigating non-iid features, it assumes that clients possess batch normalisation layers and have been actively involved in the FL training. Consequently, non-participating clients that may want to access the global model are excluded, as the method cannot generate a universally applicable global model. This fact implies limitations in distributing the global model to additional entities. Finally, as per \cite{li2022federated} experimental study, none of the existing state-of-the-art FL methods and aggregation strategies for non-iid data outperform the other ones in all cases. Therefore, exploring novel techniques to address the impact of data heterogeneity in terms of non-iid features, particularly within FL-based IDS, which is still immature in the context of the mentioned challenge, is an interesting and promising topic. To the best of our knowledge, the issue of non-iid features among clients in FL-based IDS has received limited attention.

\subsection{Contribution}
\label{Contribution}

Based on the aforementioned remarks, in this paper, we introduce the \texttt{StatAvg} method to circumvent the challenges of non-iid features in the scope of FL-driven intrusion detection. The contributions of our work are summarised as follows:
\begin{itemize}
 \item The \texttt{StatAvg} method is proposed to alleviate the effects of non-iid feature distributions in FL. According to \texttt{StatAvg}, the FL clients calculate their local data statistics, specifically the mean and variance, and transmit them to the server. The server aggregates the clients' local statistics to generate global statistics. We prove mathematically that the aggregated global statistics represent the true mean and variance of the combined datasets across all clients. Afterwards, the server broadcasts the global statistics to all clients, normalising their input features based on these global statistics.
 \item The global statistics produced by \texttt{StatAvg} can be interpreted as a universal normalisation process that can be applied along with the global model. It is important to emphasise that typically, a trained model should be accompanied by the corresponding normalisation technique on the input data. Otherwise, the model will be ineffective during inference. However, this aspect is often overlooked in the existing literature. Therefore, \texttt{StatAvg} serves as a means to offer a global normalisation technique that can be applied to the global model by external entities that are not necessarily involved in the training procedure.
 \item The performance of \texttt{StatAvg} is evaluated through experiments on open datasets for intrusion detection. Various FL aggregation strategies are used as baseline methods for comparison. The demonstrated results showcase the prevalence of \texttt{StatAvg} over the baselines in terms of evaluation metrics such as the detection accuracy and the F1 score. Finally, some illustrative insights are provided that justify the presence of clients' non-iid features on the examined intrusion detection datasets.
\end{itemize}

\section{Preliminaries of Federated Learning}
\label{Preliminaries of Federated Learning}

We consider an FL environment consisting of $N$ clients, indexed as $i \in \mathcal{N}=\{1,2,...,N\}$ and a server. Each client owns a dataset $\mathcal{D}_i=\{(\boldsymbol{x}^j_{i},y^j_i)\in\mathbb{R}^S\times\mathbb{C}\}^{D_i}_{j=1}$, where $\boldsymbol{x}^j_{i}$ is the $j$-th input sample, $D_i=\vert \mathcal{D}_i \vert$ is the number of samples and $S$ denotes the number of features. Additionally, we denote $\mathbb{C}$ as the set to which the label $y^j_i$ belongs, e.g., it could be a subset of the real numbers, a set of categorical values for classification tasks, etc. In this paper, $\mathbb{C}$ contains the labels of cyberattacks and will be described below in this work, along with the description of the datasets used in the evaluation experiments.

The overall dataset across all clients is denoted as $\mathcal{D}=\underset{i \in \mathcal{N}}{\cup }\mathcal{D}_i$ and the size of all training data is $D=\sum_{n=i}^{N}D_i$. The loss function of client $i$, is defined as:

\begin{equation}
F_i(\boldsymbol{w}) \triangleq \frac{1}{D_i}\sum_{j=1}^{D_i}\phi\left(\boldsymbol{w},\boldsymbol{x}^j_{i},y^j_i\right), \quad \forall i \in \mathcal{N},
\end{equation}
where $\phi(\boldsymbol{w},\boldsymbol{x}^j_{i},y^j_i)$ captures the error of model parameter
$\boldsymbol{w}$ for the input-output pair $(\boldsymbol{x}^j_{i},y^j_i)$. The ultimate goal of the FL process is to obtain the global parameter $\boldsymbol{w}$, which minimises the loss function on the whole dataset.
\begin{equation}
 F(\boldsymbol{w})=\sum_{n=1}^{N}n_iF_i(\boldsymbol{w}),
\end{equation}
where $n_i=\frac{D_i}{D}$ is the proportion of data samples owned by client $i$ relative to the entire dataset. 

In a nutshell, the FL process is executed for a specified number of communication rounds. At the $t$-th round, the server firstly broadcasts the global model $\boldsymbol{w}^{(t)}$ to all clients. Each client $i$ updates its local model $\boldsymbol{w}^{(t)}_i$ via a gradient-based method on the loss function $F_i$ and uploads it to the server. Finally, the server generates the global model $\boldsymbol{w}^{(t+1)}$ by using an aggregation strategy of its preference. The aforementioned process is repeated for the selected number of rounds until the convergence of the global model is achieved.

Aligning with the definitions of \cite{kairouz2021} and \cite{li2021fedbn}, the occurrence of non-iid features between clients is related to the following concepts:

\begin{itemize}
\item \textit{Feature distribution skew} (covariate shift): The marginal distributions $P_i(\boldsymbol{x})$ varies across clients, even if $P_i(y|\boldsymbol{x})$ is the same for all clients.
\item \textit{Same label, different features} (concept drift): The conditional distributions $P_i(\boldsymbol{x}|y)$ may vary across clients even if $P_i(y)$ is common. As such, the same label $y$ can have different features $\boldsymbol{x}$ for different clients.
\end{itemize}

Non-iid features may diminish the performance of FL, leading to unstable training or even divergence of the global model. This occurs because each local model is trained on a different input distribution, imposing challenges during the server's aggregation step.

\section{StatAvg - Statistical Averaging}
\label{StatAvg - Statistical Averaging}

Traditionally, individual FL clients normalise their local data based on their own local statistics, with the most prominent normalisation technique being the z-score normalisation, i.e., clients subtract the mean from each data sample of a given feature and then divide it with the standard deviation. This is equivalent to shifting the input feature distribution to have a zero mean and unit variance. Accordingly, in the testing phase, the testing dataset is scaled based on the aforementioned normalisation, individually per client. In the presence of non-iid features between clients, the local normalisation process may significantly differ from client to client. As a result, this variability may affect the convergence of the global FL model since each local model is trained on a different input data distribution. To tackle the issue of non-iid features across clients, our objective is to discover global statistics that clients can share without requiring access to their raw data. Typical statistical metrics include the mean and variance of the features, whereas this study investigates the impact of these particular metrics.

\par In the light of the previous discussion, we proceed to compute the mean and variance for each client's features.
The mean value across all samples of a feature $s \in \mathcal{S}$ of client $i$, where $\mathcal{S}$ is the entire feature set, is given as:
\begin{equation}
\label{eq:mu}
\mu_{i,s}=\frac{1}{D_i}\sum^{D_i}_{j=1}x^j_{i,s}
\end{equation}
and $\boldsymbol{\mu}_i=(\mu_{i,1},\mu_{i,2},...,\mu_{i,S})$ is the vector with all the means of each feature. Its is worth noting that $x^j_{i,s}$ is the $s$-th entry of $\boldsymbol{x}^j_{i}$. Accordingly, the corresponding variance is calculated as:
\begin{equation}
\label{eq:sigma}
\sigma^2_{i,s}=\frac{1}{D_i}\sum^{D_i}_{j=1}\left(x^j_{i,s}-\mu_{i,s}\right)^2
\end{equation}
and $\boldsymbol{\sigma}^2_i=(\sigma^2_{i,1},\sigma^2_{i,2},...,\sigma^2_{i,S})$. Hereinafter, with the term \textit{local statistics} of client $i$, we refer to the tuple $\{\boldsymbol{\mu}_i,\boldsymbol{\sigma}^2_i\}$. The \texttt{StatAvg} strategy aims at obtaining the global statistics $\{\boldsymbol{\mu}_{\mathrm{G}},\boldsymbol{\sigma}^2_{\mathrm{G}}\}$ of the overall dataset $\mathcal{D}$ by aggregating the local statistics $\{\boldsymbol{\mu}_i,\boldsymbol{\sigma}^2_i\}_{i\in\mathcal{N}}$. In this manner, all clients can normalise their data based on global statistics, which guarantees a common normalisation/scaling of the input data. The detailed process of \texttt{StatAvg} is described in Algorithm 1.

\begin{algorithm}
\caption{\texttt{StatAvg} }\label{alg:fl1}
\begin{algorithmic}[1]
\If{$t=0$}
\For{each client $i \in \mathcal{N}$} 
\State calculate $\boldsymbol{\mu}_i, \boldsymbol{\sigma}^2_i$ according to \eqref{eq:mu}, \eqref{eq:sigma}
\State send $\boldsymbol{\mu}_i, \boldsymbol{\sigma}^2_i$ to the server
\EndFor
\State server calculates the global statistics as:

\hspace{3em}$\boldsymbol{\mu}_{\mathrm{G}}=\sum_{i\in\mathcal{N}}n_i\boldsymbol{\mu}_i,$
\vspace{0.2cm}

\hspace{1em}$\boldsymbol{\sigma}^2_{\mathrm{G}}= \sum_{i\in\mathcal{N}}n_i\left(\boldsymbol{\sigma}^2_i+(\boldsymbol{\mu}_i-\boldsymbol{\mu}_{\mathrm{G}})^2\right)$
\vspace{0.2cm}
\State server sends $\boldsymbol{\mu}_{\mathrm{G}},\boldsymbol{\sigma}^2_{\mathrm{G}}$ to all clients
\For{each client $i \in \mathcal{N}$} 
\State normalise input features as:
\vspace{0.2cm}

\hspace{2em}$\tilde{\boldsymbol{x}}^j_{i,s}=\frac{\boldsymbol{x}^j_{i,s}-\mu_{\mathrm{G},s}}{\sigma_{\mathrm{G},s}}, \quad \forall j\in\{1,...,D_i\},\,\,\forall s \in \mathcal{S}$
\EndFor
\Else{ execute conventional FL process}
\EndIf
\end{algorithmic}
\end{algorithm}

As can be seen, the \texttt{StatAvg} strategy occurs solely during the first round ($t=0$), prior to the actual FL training. Firstly, in steps 2 - 4, each client calculates its local statistics and sends them to the server. Following that, in steps 5 - 6, the server calculates the global statistics based on the received local statistics and broadcasts them back to the clients. It is worth mentioning that the operations in step 5 are carried out element-wise. The rationale behind the aggregation technique used to obtain $\boldsymbol{\mu}_\mathrm{G}$ and $\boldsymbol{\sigma}^2_\mathrm{G}$ is explained later in this work. Afterwards, in steps 7 - 8, the clients normalise their input features based on the global statistics by utilising conventional z-score normalisation. It should be highlighted that the communication overhead for exchanging the local and global statistics between the clients and the server is negligible since it takes place solely during the first round. Finally, at step 9, a conventional FL process follows, e.g., the FedAvg. The selection of the aggregation strategy at step 9 can vary according to the particularities of the underlying FL task. 

\begin{figure*}[t!]
 \centering \includegraphics[width=0.85\linewidth]{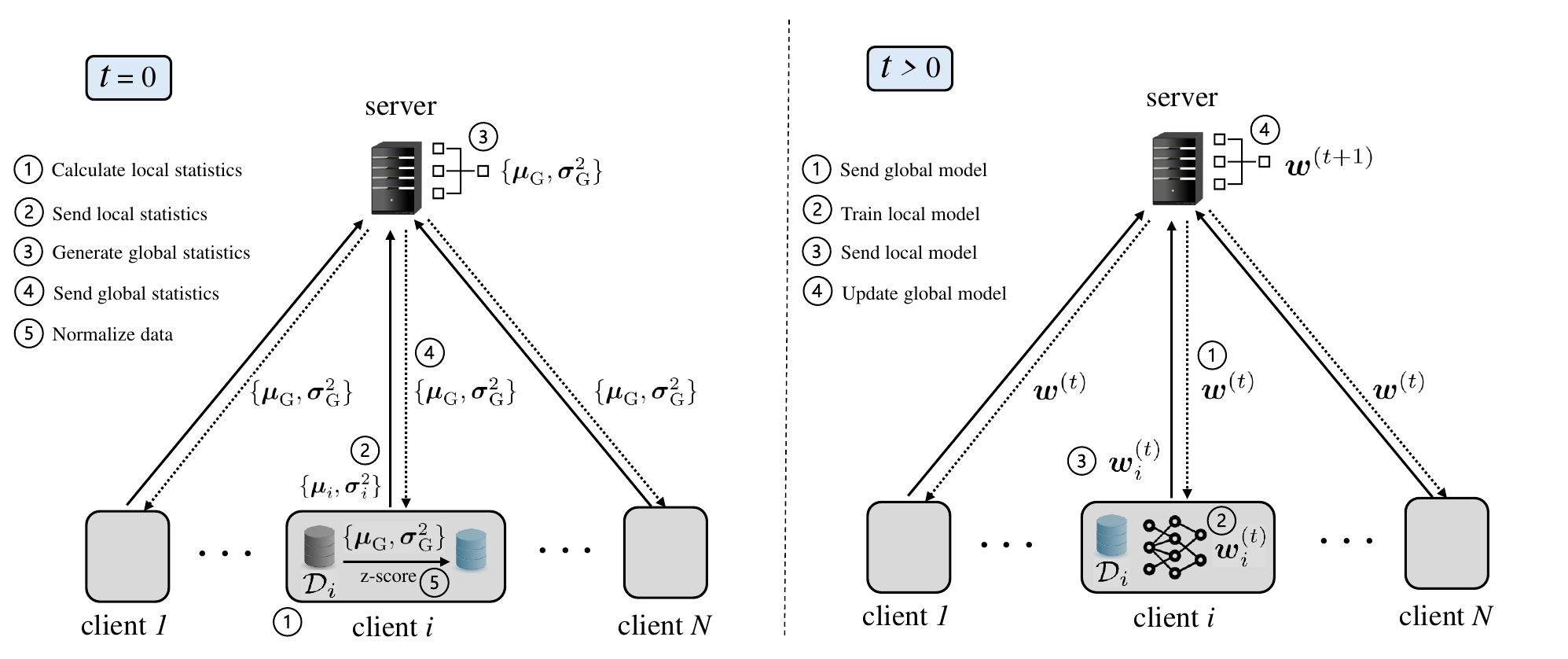}
 \caption{Visual representation of StatAvg design and implementation.}
 \label{fig:arch}
\end{figure*}

It should be again clarified that \texttt{StatAvg} focuses on the aggregation of statistical metrics rather than local models $\boldsymbol{w}^{(t)}_i$, facilitating its integration with any model aggregation strategy. Fig.~\ref{fig:arch} provides an illustration of \texttt{StatAvg}'s implementation. Finally, we stress that through \texttt{StatAvg}, a universal input data normalisation technique is provided. This is a crucial remark since a trained model should be paired with the appropriate data normalisation technique to render it effective during inference.

In the continue, we will show that $\boldsymbol{\mu}_\mathrm{G}$ and $\boldsymbol{\sigma}^2_\mathrm{G}$ are the mean and variance of the overall dataset $\mathcal{D}$. First, we assume that $\mathcal{D}_i\cap\mathcal{D}_k=\emptyset, \, \forall i,k \in \mathcal{N}, \, i\neq k$. This implies that all local datasets are pairwise disjoint. The assumption is reasonable, considering that each dataset originates from a distinct client, thus making it highly unlikely - if not impossible - for identical samples to appear across different local datasets. To this end, we proceed to formulate the following proposition.

\begin{proposition}
 Let $\boldsymbol{x}_{i,s}\in\mathbb{R}^{D_i}$ be the vector containing the $s$-th feature across all samples of $\mathcal{D}_i$. Also, let $\boldsymbol{z}_s=(\boldsymbol{x}_{1,s},...,\boldsymbol{x}_{N,s})$ be the concatenation of all clients vectors, with $\boldsymbol{z}_s\in\mathbb{R}^D$. The mean and variance of $\boldsymbol{z}_s$ are given as
 \begin{equation}
 \begin{split}
\mu_{\mathrm{G},s} &=\sum_{i\in\mathcal{N}}n_i\mu_{i,s}\\
\sigma^2_{\mathrm{G},s}&= \sum_{i\in\mathcal{N}}n_i\left(\sigma^2_{i,s}+(\mu_{i,s}-\mu_{\mathrm{G},s})^2\right).
 \end{split}
 \end{equation}
\end{proposition}
\begin{IEEEproof}

First, the notation of $s$ is dropped for the simplicity of presentation. It is straightforward to compute the mean of $\boldsymbol{z}$ as:

\begin{equation}
\begin{split}
 \mu_{\mathrm{G}}&=\frac{1}{D}\sum_{l=1}^{D}z_l=\frac{1}{D}\sum_{i=1}^{N}\sum_{j=1}^{D_i}x^{j}_i=\frac{1}{D}\sum_{i=1}^{N}D_i\sum_{j=1}^{D_i}\frac{x^{j}_i}{D_i}\\
 &=\sum_{i=1}^{N}n_i\mu_i.
\end{split}
\end{equation}

Before examining $\sigma^2_{\mathrm{G}}$, first, it is noted that for the local variances, it holds:

\begin{equation}
 \sigma^2_in_i=\sum_{j=1}^{D_i}(x^j_i-\mu_i)^2
, \quad \forall i \in \mathcal{N}. 
\end{equation}

Similarly, for $\boldsymbol{z}$ we get
\begin{equation}
\label{eq:inner}
\begin{split}
 \sigma^2_{\mathrm{G}}D&=\sum_{l=1}^{D}(z_l-\mu_{\mathrm{G}})^2=\sum_{i=1}^{N}\sum_{j=1}^{D_i}(x^{j}_i-\mu_{\mathrm{G}})^2\\
 &=\sum_{i=1}^{N}\sum_{j=1}^{D_i}\left((x^{j}_i)^2-2x^{j}_i\mu_{\mathrm{G}}+\mu^2_{\mathrm{G}}\right).
\end{split}
\end{equation}

The inner sum in the last term of \eqref{eq:inner} can be expanded by adding and subtracting $\mu^2_i$, as:

\begin{equation}
\label{eq:long}
 \begin{split}
&\sum_{j=1}^{D_i}\left((x^{j}_i)^2-2x^{j}_i\mu_{\mathrm{G}}+\mu^2_{\mathrm{G}}+\mu^2_i-\mu^2_i\right)\\
&=\sum_{j=1}^{D_i}\left((x^{j}_i-\mu_i)^2+2x^{j}_i(\mu_i-\mu_{\mathrm{G}})+\mu^2_{\mathrm{G}}-\mu^2_i\right)\\
&=\sum_{j=1}^{D_i}(x^{j}_i-\mu_i)^2+2D_i\mu_i(\mu_i-\mu_{\mathrm{G}})+D_i\mu^2_{\mathrm{G}}-D_i\mu^2_i\\
&=D_i\sigma^2_i + D_i\mu^2_{\mathrm{G}} - 2D_i\mu_i\mu_{\mathrm{G}} + D_i\mu^2_i\\
&=D_i\sigma^2_i + D_i(\mu_i-\mu_{\mathrm{G}})^2.
 \end{split}
\end{equation}

By combining \eqref{eq:long} with \eqref{eq:inner} we conclude to
\begin{equation}
 \sigma^2_{\mathrm{G}}= \sum_{i\in\mathcal{N}}n_i\left(\sigma^2_i+(\mu_i-\mu_{\mathrm{G}})^2\right),
\end{equation}
which completes the proof.
\end{IEEEproof}

Proposition 1 provides a way to obtain the global mean and variance across the whole dataset $\mathcal{D}$ for a given feature $s$. The proof can be easily generalised $\forall s \in \mathcal{S}$, which gives rise to the vector representation of the global mean and variance for each feature, i.e., $\boldsymbol{\mu}_{\mathrm{G}}$ and $\boldsymbol{\sigma}^2_{\mathrm{G}}$, respectively. This result is used in step 5 of Algorithm 1 to derive the global mean and variance.

\section{Evaluation Analysis}
\label{Evaluation Analysis}

\begin{table*}[t!]
\centering
\caption{List of notations}
\begin{tabular}{P{1.5cm}|P{4cm}|P{1.5cm}|P{7cm}}
\hline 
\textbf{Parameter}& \textbf{Description} & \textbf{Parameter} & \textbf{Description} \\
\hline \hline
$\mathcal{N}$ & Set of clients & $P(\cdot)$ & Probability density function \\
\hline
$i$ & Indexing of clients & $\boldsymbol{\mu}_i$ & The mean values vector of the features for the $i$-th client \\
\hline
$\mathcal{D}_i$ & Dataset of client $i$ &$\boldsymbol{\sigma}^2_i$ & The variances vector of the features for the $i$-th client \\
\hline
$D_i$ & Dataset size of client $i$ & $\boldsymbol{\mu}_{\mathrm{G}}$ & Global mean values vector \\ \hline
$\boldsymbol{x}^j_i$ & $j$-th input sample of client $i$ & $\boldsymbol{\sigma}^2_{\mathrm{G}}$ & Global variances vector\\ \hline
$S$ & Number of features & $\boldsymbol{z}$ & Concatenated vector \\ \hline
$s$ & Index of features & $z_l$ & The $l$-th element of the vector $\boldsymbol{z}$ \\ \hline
$y^j_i$ & $j$-th label of client $i$ & $t$ & FL round index\\ \hline
$\mathcal{D}$ & Overall dataset of all clients & $n_i$ & Proportion of data samples owned by client $i$\\ \hline
$D$ & Size of the overall dataset & $\boldsymbol{w}^{(t)}$ & Global model at round $t$ \\
\hline
$N$ & Total number of clients & $\boldsymbol{w}^{(t)}_i$ & Local model of $i$-th client at round $t$ \\
\hline 
\end{tabular}
\end{table*}

This section presents experiments conducted on different datasets to detect intrusions in a federated setting. The effectiveness of the proposed strategy \texttt{StatAvg} is evaluated by comparing it with various baseline methods.

\subsection{Evaluation Datasets}
\label{Evaluation Datasets}

The experiments were conducted on the following well-known public datasets.\\

\noindent \textbf{TON-IoT Dataset}: Among others, the TON-IoT Dataset \cite{ton} includes operating system data of Ubuntu versions 14 and 18, which is adopted in our work. More specifically, it includes audit traces documenting memory activities within the operating system. The dataset is suitable for training and designing host-based IDS. Also, the dataset is composed of data stemming from various physical or virtual devices belonging to the edge and cloud layers. The description of the selected features is provided in Table~\ref{table:ton_features}. Furthermore, the attacks on the host system that serve as the labels of the dataset are ``dDoS'', ``DoS'', ``Injection'', ``Password'', ``Mitm'', while also a class named ``Normal'' is included, indicating the normal behaviour of the host system. More details regarding the dataset can be found in \cite{ton} and \cite{moustafa2020}. \\

\noindent \textbf{CIC-IoT-2023 Dataset}: The CIC-IoT-2023 Dataset \cite{neto2023} is a realistic IoT attack dataset, using an extensive topology composed of multiple IoT devices designated as either attackers or targets. The dataset entails 48 features that are characterised by metrics such as packet flow statistics, employed application layer protocols, TCP flags, etc. As we do not explicitly describe all features for brevity, additional information can be found in \cite{neto2023}. Furthermore, the dataset categorises attacks into eight classes, namely ``Brute force'', ``dDoS'', ``DoS'', ``Mirai'', ``Recon'', ``Spoofing'', ``Web-based'', and ``Normal''.

\begin{table}[t!]
\centering
\caption{TON-IoT Dataset: Feature Description}
\label{table:ton_features}
\begin{tabular}{P{1.2cm}|P{6.2cm}}
\hline 
\textbf{Feature name}& \textbf{Description} \\
\hline \hline
MINFLT & The number of page faults issued by this process that have been solved by
reclaiming the requested memory page from the free list of pages \\
\hline
MAJFLT & The number of page faults issued by this process that have been solved by
creating/loading the requested memory page \\
\hline
VSTEXT & The virtual memory size used by the shared text of this process\\
\hline
VSIZE & The total virtual memory usage consumed by this process \\
\hline
RSIZE & The total resident memory usage consumed by this process \\
\hline
VGROW & The amount of virtual memory that the process has grown during the last
interval \\
\hline
RGROW & The amount of resident memory that the process has grown during the last
interval \\
\hline
MEM & Memory occupation percentage \\
\hline
\end{tabular}
\end{table} 

\subsection{Baseline Aggregation Methods}
\label{Baseline Aggregation Methods}

To evaluate the performance of \texttt{StatAvg}, we use the following baseline aggregation strategies:

\noindent \textbf{FedAvg}: It is the de facto approach for FL \cite{mcmahan2017}. Clients perform local model updates and the server executes the aggregation of the local models to generate the global model.\\

\noindent \textbf{FedLN}: The layer normalisation is included in the local models for mitigating the effects of non-iid features \cite{du2022}. FedLN performs local updates and averages local models similarly to FedAvg. \\

\noindent\textbf{FedBN}: Employs local batch normalisation (BN) to the local models prior to averaging them towards alleviating feature shift. Nonetheless, FedBN assumes that local models have BN layers and omits their parameters from the aggregation step at the side of the server \cite{li2021fedbn}.

It is worth noting that FedLN and FedBN are specially tailored to address the issue of non-iid features, justifying their selection. As follows, in FedAvg, FedLN and FedBN, the normalisation of the local training data is performed based on the local client statistics, aligning with the conventional FL approach. Also, the testing data undergo scaling in accordance with the respective local normalisation for each client individually. Finally, it is noted that the proposed technique \texttt{StatAvg} utilises FedAvg at step 9 of Algorithm 1 as the default model aggregation strategy.

\subsection{Experimental Setup}
\label{Experimental Setup}

The following settings apply to all experiments unless specified otherwise. The number of clients has been set as $N=5$, and all clients are considered to participate in every FL round. Also, the number of FL rounds is set to 50. Each client receives an equal proportion $n_i=\frac{1}{N}$ of the original dataset $\mathcal{D}$. Moreover, the division is conducted through stratification based on the labels of the original dataset, aiming to approximate a common $P_i(y)$ across all clients. This implies that clients share a common label distribution. Following that, each client splits its local dataset into training and testing subsets, with a ratio of 4 to 1. Due to the significant class imbalance in the datasets, each client generates synthetic instances from the minority classes in the training set by using SMOTE \cite{chawla2002}. 

The local model of each client is a neural network consisting of 3 Fully Connected (FC) hidden layers with 128 neurons and ReLU activation, denoted as (FC(128), ReLU), followed by a softmax activation on the output layer. In the case of the baselines FedLN and FedBN, layer normalisation (LN) and batch normalisation (BN) layers are incorporated into the local models, resulting in each layer being structured as (FC(128), ReLU, LN) and (FC(128), BN, ReLU), respectively. For the local training updates, the Adam optimiser is adopted \cite{kingma2014adam}. Finally, additional settings are summarised in Table \ref{table:settings}.

\begin{table*}[t!]
\centering
\caption{Experimental Settings}
\label{table:settings}
\begin{tabular}{P{6cm}|*{2}{P{1.8cm}}}
\hline 
Datasets & TON-IoT & CIC-IoT-2023 \\
\hline \hline
 training samples per client & 19616 & 64050 \\
\hline
training samples per client (SMOTE upsampling) & 84000 & 373176 \\
\hline
local training epochs & 2 & 1 \\
\hline
batch size & 512 & 1024 \\
\hline
learning rate & 0.002 & 0.01 \\
\hline
\end{tabular}
\end{table*}

\subsection{Evaluation Results}
\label{Evaluation Results}

Regarding the performance evaluation, we use common evaluation metrics such as the confusion matrix, accuracy, and F1 score. Given a specific attack/class, the confusion matrix includes the following standard metrics: the True Positive (TP) represents instances where the model correctly identifies a sample as belonging to a specific attack type. True Negative (TN) counts instances where the model accurately identifies a sample as not belonging to a specific attack type when it truly does not. False Positive (FP) denotes instances where the sample is predicted as of a certain attack, but actually, the sample does not belong to that attack type. False Negative (FN) is the number of instances for which the model fails to predict a sample as a specific attack type, even though the sample actually belongs to that attack. Next, the accuracy and F1 score are defined as:

\begin{equation}
\mathrm{ACC}=\frac{\mathrm{TP}+\mathrm{TN}}{\mathrm{TP}+\mathrm{TN}+\mathrm{FP}+\mathrm{FN}}
\end{equation}
and
\begin{equation}
\mathrm{F1}=\frac{2\mathrm{TP}}{2\mathrm{TP}+\mathrm{FP}+\mathrm{FN}},
\end{equation}
respectively. 

The evaluation metrics showcased in the results have been averaged across all classes due to the multi-class nature of the problems we are addressing. Finally, it is noted the evaluation was performed using clients' testing sets, and the demonstrated results were also averaged across all clients.

\begin{figure}[httb!]
\label{fig:acc_ton}
 \centering \includegraphics[width=1\linewidth]{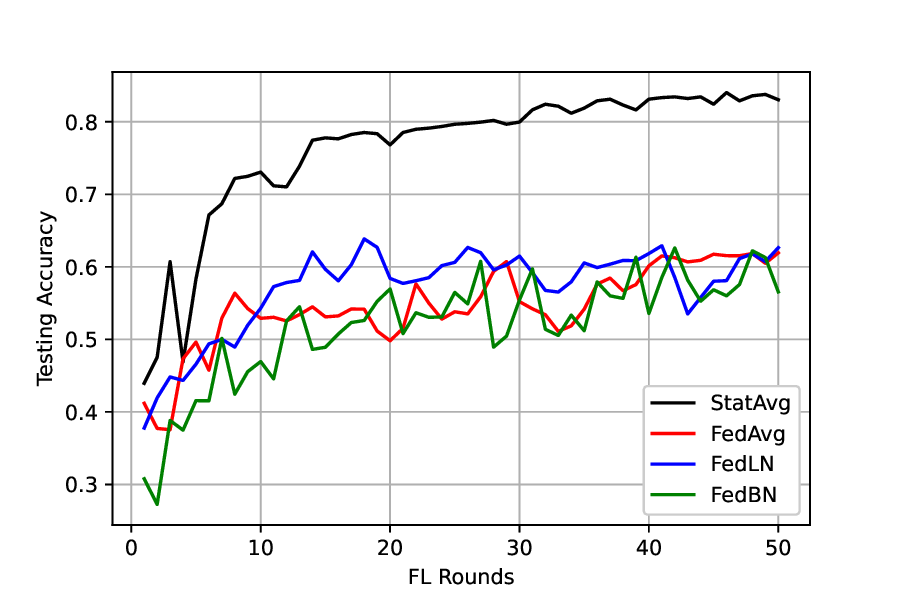}
 \caption{Testing accuracy on TON-IoT dataset.}
 \label{fig:Acc_linux}
\end{figure}

\begin{figure}[httb!]
 \centering \includegraphics[width=1\linewidth]{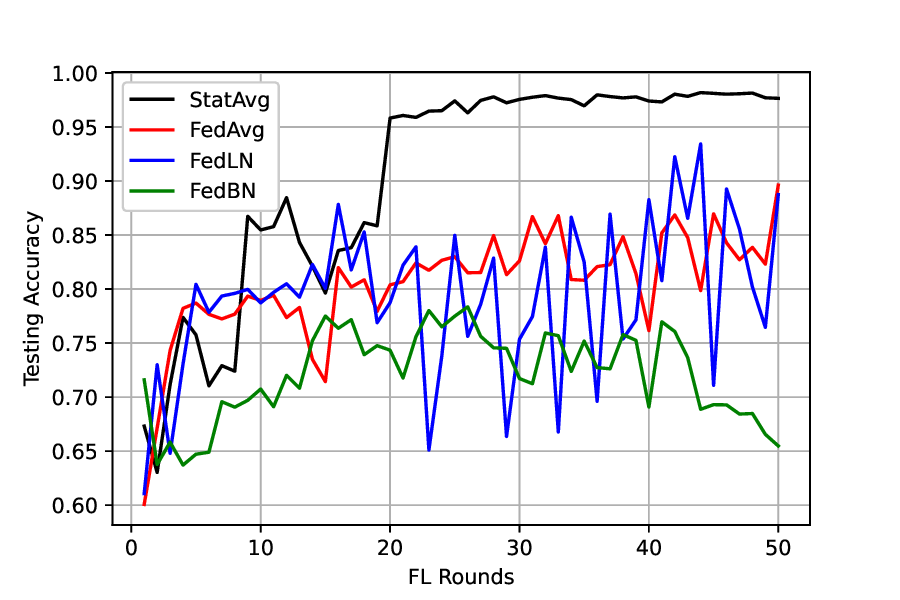}
 \caption{Testing accuracy on CIC-IoT-2023 dataset.}
 \label{fig:Acc_cic}
\end{figure}

First, the evolution of testing accuracy throughout the FL rounds is evaluated. In Fig. \ref{fig:Acc_linux} and Fig. \ref{fig:Acc_cic}, the \texttt{StatAvg} strategy is compared with the selected baselines on the TON-IoT and CIC-IoT-2023 datasets, respectively. It is evident that \texttt{StatAvg} significantly outperforms the baseline strategies across both datasets in terms of accuracy. Moreover, the convergence curve of \texttt{StatAvg} is more stable compared to that of the baseline methods, which display higher variance. The exhibited performance gain lies in the fact that \texttt{StatAvg} utilises global statistics to normalise the clients' features. Although FedLN and FedBN have been designed to minimise the effects of non-IID features between clients, it is discernible that they struggle to address this issue in certain datasets. The variations in local client statistics, and consequently, the diverse local normalisation utilised, appear to degrade the performance of FL.

\begin{figure}[httb!]
 \centering \includegraphics[width=0.90\linewidth]{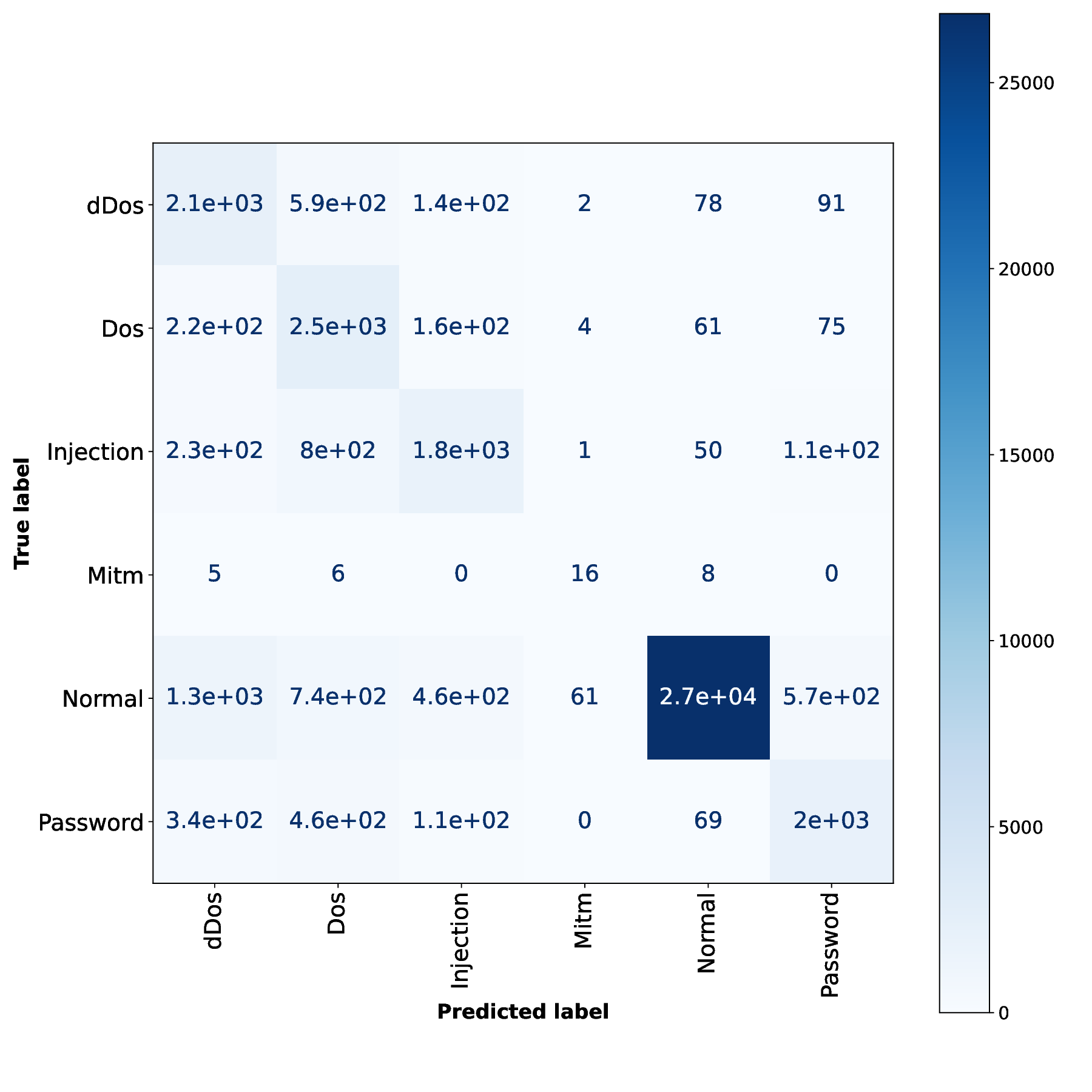}
 \caption{Confusion matrix of StatAvg on TON-IoT dataset.}
 \label{fig:cnf_linux}
\end{figure}

\begin{figure}[ht!]
 \centering \includegraphics[width=0.90\linewidth]{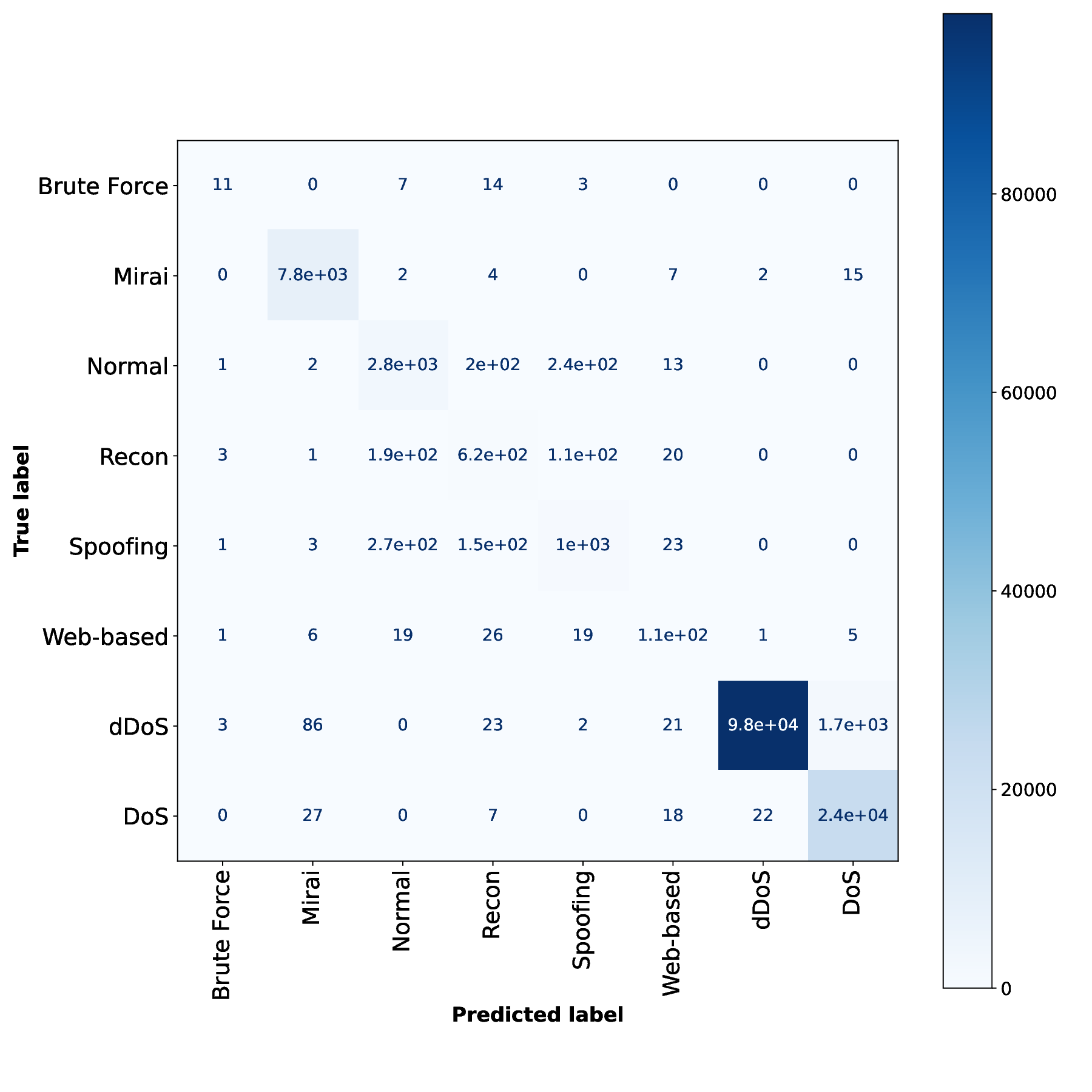}
 \caption{Confusion matrix of StatAvg on CIC-IoT-2023 dataset.}
 \label{fig:cnf_cic}
\end{figure}

Moreover, in Table~\ref{tab:eval_ton} and Table~\ref{tab:eval_cic}, some evaluation metrics for the case of TON-IoT and CIC-IoT-2023 datasets are demonstrated, respectively. The considered metrics showcase the performance of the best models encountered during the FL training for each strategy. It can be observed that \texttt{StatAvg} has superior performance against the baseline strategies. Specifically, in the case of the TON-IoT dataset, \texttt{StatAvg} demonstrates a notable improvement of over 19\% and 21\% in accuracy and F1 score, respectively, compared to the second-best strategy FedLN. Also, when considering the CIC-IoT-2023 dataset, the corresponding increase is over 4\% and 2\% for accuracy and F1 score. The detailed confusion matrices of the \texttt{StatAvg} method are presented in Fig.~\ref{fig:cnf_linux} and Fig. \ref{fig:cnf_cic}, for the TON-IoT and CIC-IoT-2023 datasets, respectively.

\begin{table}[httb!]
\centering
\caption{Evaluation Metrics on TON-IoT Dataset}
\label{tab:eval_ton}
\begin{tabular}{P{1.8cm}|*{4}{P{0.9cm}}}
\hline 
Strategy & ACC & TPR & FPR & F1 \\
\hline \hline
\textbf{StatAvg} & \textbf{83.93}\% & \textbf{69.26}\% & \textbf{3.13}\% & \textbf{62.36}\% \\
\hline
FedAvg & 63.68\% & 48.7\% & 8.22\% & 38.30\% \\
\hline
FedLN & 64.29\% & 48.9\% & 7.69\% & 40.73\% \\
\hline
FedBN & 62.60\% & 46.85\% & 8.66\% & 36.99\% \\
\hline
\end{tabular}
\end{table}

\begin{table}[httb!]
\centering
\caption{Evaluation Metrics on CIC-IoT-2023 Dataset}
\label{tab:eval_cic}
\begin{tabular}{P{1.8cm}|*{4}{P{0.9cm}}}
\hline 
Strategy & ACC & TPR & FPR & F1 \\
\hline \hline
StatAvg & \textbf{97.64}\% & \textbf{76.01}\% & \textbf{0.33}\% & \textbf{75.63}\% \\
\hline
FedAvg & 89.71\% & 70.16\% & 3.34\% & 69.82\% \\
\hline
FedLN & 93.42\% & 73.41\% & 1.58\% & 73.59\% \\
\hline
FedBN & 78.34\% & 69.33\% & 4.10\% & 66.32\% \\
\hline
\end{tabular}
\end{table}

\begin{figure}[httb!]
 \centering
 \subfigure[Unprocessed dataset.]{\includegraphics[keepaspectratio,width=0.49\linewidth]{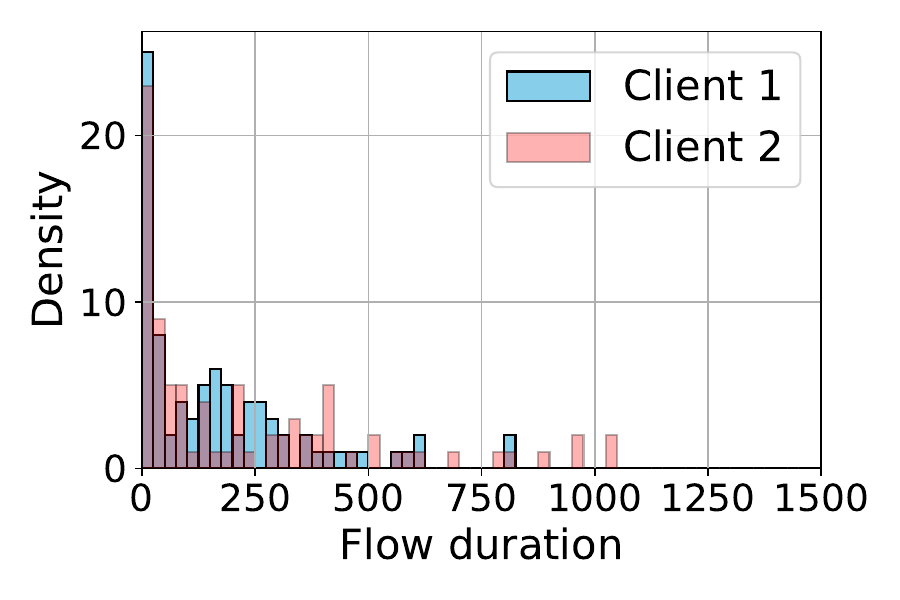}}
 \subfigure[Upsampled (SMOTE) dataset.]{\includegraphics[keepaspectratio,width=0.49\linewidth]{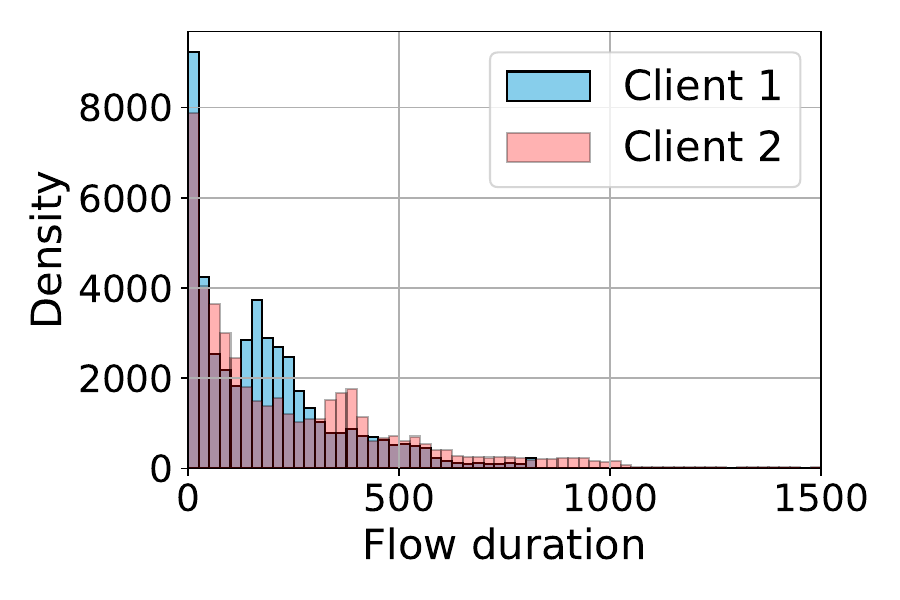}}
 \caption{Distribution of the feature ``Flow Duration", given the attack label ``Web-based", on CIC-IoT-2023 dataset.}
 \label{distributions}
\end{figure}

To shed light on the concept of non-iid features, we present some illustrative examples derived from the examined datasets. First, we take a deeper look into the training samples of the CIC-IoT-2023 dataset, focusing specifically on those labelled with the attack category $y=\text{``Web-based''}$. Fig.~\ref{distributions} illustrates the distribution of the feature ``Flow Duration'' for the clients $i=\{1,2\}$, formally written as $P_i(x_{i,s}|y=\text{``Web-based''})$ where $s=\text{``Flow Duration''}$. It can be observed from Fig.~\ref{distributions} (a) that the distributions of the clients differ. Nevertheless, it remains uncertain whether this disparity in distributions is inherent or if it is related to the limited number of samples within the selected class. It is worth noting that the "Web-based" class is indeed a minority class. From Fig.~\ref{distributions} (b), it is evident that the difference in distributions persists after upsampling the dataset via SMOTE. This example shows that even if $P_i(y)$ is approximately the same for all clients, as previously explained in the experimental setup, the conditional distributions $P_i(\boldsymbol{x}|y)$ can still differ. This phenomenon is related to the concept of \textit{Same label, different features}, discussed in Section~\ref{Related Work, Motivation and Contributions}. Another example that highlights the differences in the distributions of features is presented in Table~\ref{tab:client_metrics}. Here, statistical metrics for selected features from the TON-IoT dataset have been calculated. It can be observed that the feature ``VSIZE'' demonstrates consistent mean and variance across clients, while the feature ``MINFLT'' displays high variations in the statistical metrics. This example highlights the statistical differences that some features may exhibit among clients, which in turn influences the local normalisation of the features and potentially hinders the FL stability and convergence.

\begin{table}[htbp]
 \centering
 \caption{Statistical Metrics of Clients' Features on TON-IoT Dataset}
 \label{tab:client_metrics}
 \begin{tabular}{cccccc}
 \toprule
 \textbf{Feature name} & \multicolumn{2}{c}{MINFLT} & \multicolumn{2}{c}{VSIZE} \\
 \cmidrule(lr){2-3} \cmidrule(lr){4-5}
 & Mean & Variance & Mean & Variance \\
 \midrule
 Client 1 & $694.1$ & $5.8\cdot10^6$ & $8621.3$ & $1.32\cdot10^8$ \\
 Client 2 & $694.8$ & $3.5\cdot10^6$ & $8663.3$ & $1.33\cdot10^8$ \\
 Client 3 & $735.7$ & $5.8\cdot10^7$ & $8364.9$ & $1.26\cdot10^8$ \\
 Client 4 & $691.3$ & $3.9\cdot10^6$ & $8521.9$ & $1.31\cdot10^8$ \\
 Client 5 & $769.8$ & $1.8\cdot10^8$ & $8519.1$ & $1.3\cdot10^8$ \\
 \bottomrule
 \end{tabular}
\end{table}


\section{Conclusions}
\label{Conclusions}

This paper proposes the \texttt{StatAvg} technique for mitigating the impact of non-iid features among clients in FL settings. The key aspect of \texttt{StatAvg} is to produce global data statistics based on the local data statistics of FL clients. The generation of global statistics, which is carried out by the server, gives rise to a universal data normalisation technique that is performed by all clients. Particular attention is given to FL-based IDS, which is the focus of the experiments that were conducted. The results corroborate the effectiveness of \texttt{StatAvg} in providing robust FL convergence and classifying cyber-attacks compared to various baseline FL schemes. Moreover, valuable insights are offered within the scope of non-iid features among clients for the selected intrusion detection datasets. Finally, as \texttt{StatAvg} precedes the actual FL procedure, it can be combined with any FL aggregation strategy, a topic which is left for future investigation. Moreover, the applicability of \texttt{StatAvg} is not limited solely to FL-based IDS, as its efficacy may encompass any FL application associated with non-iid features among clients.

\bibliographystyle{IEEEtran}
\bibliography{main}

\end{document}